\documentstyle[preprint,eqsecnum,aps]{revtex}

\tightenlines
\begin{document}
\draft
\preprint{}
\title{ Evolution of the low-frequency spin dynamics in ferromagnetic manganites \\}
\author{J. A. Fernandez-Baca,$^1$ P. Dai,$^1$ H. Y. Hwang,$^2$ S-W.
Cheong,$^{2,3}$ and C. Kloc$^2$}
\address{
$^1$ Oak Ridge National Laboratory, Oak Ridge,
Tennessee 37831-6393}
\address{$^2$ Bell Laboratories, Lucent Technologies, Murray Hill, NJ 07974}
\address{$^3$ Department of Physics and Astronomy, Rutgers University,
Piscataway, New Jersey 08855}

\date{\today}
\maketitle
\begin{abstract}
Elastic and inelastic neutron scattering was used to study two ferromagnetic
manganites A$_{1-x}$B$_{x}$MnO$_3$ (x $\approx$ 0.3) with $T_c$=197.9 K and  300.9 K. The spin dynamical behavior of these is similar at  low temperatures, but drastically different at temperatures around $T_c$. While 
the formation of spin clusters of size ($\sim20$ \AA) dominates the spin dynamics of the 197.9 K sample close to   $T_c$, the paramagnetic to
ferromagnetic transition for the  300.9 K sample is more conventional.
These results, combined with seemingly inconsistent earlier reports,
reveal clear systematics in the spin dynamics of the manganites.

\end{abstract}


\narrowtext
The prospect of potential technological applications of the so-called colossal magnetoresistance (CMR) 
\cite{chabara}
has lead to a revival in the study of perovskite manganites $A_{1-x}B_x$MnO$_3$
(where $A$ and $B$ are trivalent and divalent ion respectively).  For
compositions of $x \approx 0.3$, these materials exhibit a resistivity drop that is intimately related to the paramagnetic to ferromagnetic ordering at the Curie temperature ($T_C$). The magnitude of the resistivity drop and $T_C$ can be tuned
 continuously by different $A$-site substitutions \cite{hwang1}. The central issue is
whether the microscopic understanding of these systems is complete
within the double exchange (DE) model \cite{zener} or requires additionally a strong
Jahn-Teller based electron-lattice coupling \cite{millis,roder}.
Experimentally, large oxygen lattice distortions have been found in La$_{0.7}$Ca$_{0.3}$MnO$_3$ ($T_C\sim 250$ K) \cite{dai,radaelli1,billinge}, consistent with the existence of  lattice and/or magnetic 
polarons  ( 10-20 \AA \ )\cite{teresa,erwin}. These results  may also  be related to the development of a  prominent diffusive central peak close to $T_C$ in this compound \cite{lynn}.
 By contrast, no signatures of  unusual oxygen lattice distortions or   
unconventional spin dynamics  have been reported \cite{martin,toby,endoh}
in  higher $T_C$  materials. 
In particular, Perring {\it et al.} \cite{toby} found that the spin wave
dispersion throughout the Brillouin zone in a La$_{0.7}$Pb$_{0.3}$MnO$_3$
($T_C=355$ K) crystal
can be described by the conventional  Heisenberg  ferromagnet (FM)
model with only the nearest-neighbor  exchange coupling. 
Furukawa \cite{furukawa} has recently shown that this result  can be fully explained within the DE model.

Thus, from the current experimental evidence it is difficult to develop a
consistent picture of the spin dynamical behavior in the ferromagnetic perovskite manganites.  Furthermore, 
it is not clear whether the reported  polarons are the driving force of the CMR effect or  merely a consequence of the insulator-to-metal transition from the DE mechanism itself.

By judicially tuning the  lattice distortions through
different $A$-site substitutions, we prepared two high
quality  perovskite  manganite single crystals
that have transition temperatures of  197.9 K
(Nd$_{0.7}$Sr$_{0.3}$MnO$_3$, or
NSMO)  and 300.9 K (Pr$_{0.63}$Sr$_{0.37}$MnO$_3$, or PSMO).
 Our aim was to study, via elastic and
inelastic neutron scattering, the systematics of the  unique coupling
between the
lattice distortion and the spin dynamical properties of these systems. We
show that  the spin dynamical behavior of these two systems is similar at low
temperature, but drastically different for temperatures around $T_C$.  While
the spontaneous formation of spin polarons ($\sim20$ \AA \ in size) close to
$T_C$
dominates the spin dynamics of the 197.7 K sample, the paramagnetic to
ferromagnetic phase transition for the 300.9 K material is more conventional.

Large single crystals ($\approx$ 1cc) of
NSMO and PSMO of mosaic spreads of $\sim 0.5$ degree full-width at
half-maximum (FWHM) were grown by the floating zone method.
The neutron scattering experiments were carried out
 using the HB-1 triple-axis spectrometer at the
High-Flux Isotope Reactor at the Oak Ridge
National Laboratory. We have used pyrolytic graphite (PG) as monochromator
and crystals of
PG or Be as the analyzer to select neutrons with a wavelength of 2.46 \AA \
($E=13.6$ meV) or 1.64 \AA \ ($E=30.5$ meV).
The samples were slightly orthorhombic and twinned at the measured
temperatures, but for the purpose of our measurements (and the utilized
instrumental resolution) we assumed a cubic lattice ($a=3.86$ \AA) for both
crystals.

The temperature dependence of the intensity of the (0 0 1) Bragg reflection
for both samples [see Figs 1(a) and 1(b)] indicates that
ferromagnetic ordering develops at low temperatures.  These
measurements were done
allowing sufficient time ($\approx$ 20 min) for the samples to reach thermal
equilibrium and were performed both on cooling (full circles) and on
warming (open circles). There is a  small but reproducible
hysteresis in both  data sets,
suggesting that the ferromagnetic phase transition is
weakly first order. The difference between the transition
temperatures on
cooling and on warming was less than $\approx$ 0.5 K, and
for the purposes of this experiment we chose the average of these two as
the transition temperatures (197.9 $\pm$ 0.5 K for NSMO and 300.9$\pm$ 0.5 K
for PSMO). The ferromagnetic-to-paramagnetic transition
 for NSMO is significantly rounder than that for PSMO, this
may be due to the special features of the former system which will
be addressed below. We also note  that, within the precision of our measurements, there was no evidence of spin canting \cite{de gennes},  or of a spin diffusive peak at the antiferromagnetic positions in any of the two samples.

The spin-wave dispersion curve for NSMO throughout the  $ (0, 0, 1 + \zeta)$ 
Brillouin zone at  10 K is plotted in Figure 1(c) (solid circles), which also shows similar data for PSMO (open  circles)\cite{hwang2}.  Note that despite the large difference in the transition temperatures of these two systems, the two curves are remarkably similar.  This observation cannot be explained by any simple model of magnetic ordering.  In particular,   for a Heisenberg ferromagnet with only nearest neighbor  exchange interaction $J$,   $kT_C$ is nearly proportional to  $J$\cite{calculation}. To elucidate the nature of the spin dynamics of these two systems,  we performed high-resolution neutron inelastic scattering measurements.   In the long wavelength limit (small wavectors {\bf q}), the measured spin-wave energies can be fitted to the well-known  quadratic dispersion relation  $\hbar\omega=\Delta+Dq^2$, where $D$ is the spin wave stiffness constant and $\Delta$ is a small dipolar gap.  In the Heisenberg model, $D=2JSa^2$,  where $S$ is the magnitude of the electronic spin at the magnetic ion sites and $a$ is the lattice parameter. The quadratic dispersion form, however, is  general and not limited to this model.

Figures 2(a) and 2(b) show the spin-wave energies vs. $q^2$ for PSMO and NSMO at various temperatures and the fits to the above quadratic dispersion relation. In all cases a very small ($\le$ 0.05 meV) energy gap was obtained from the fits, but for practical purposes this gap was assumed to be zero. 
To follow the spin dynamical properties closer to the transition temperatures,
we show in Figure 2(c) the spin-wave stiffness constant
$D(T)$ as a function of $T/T_C$  for both samples.  As $T \rightarrow T_C$, the hydrodynamic
and the mode-mode coupling theories \cite{collins} predict a temperature
dependence of the spin-wave stiffness of the form $D(T)\sim
[(T-T_C)/T_C]^{\nu - \beta}$.  The fit of the spin-wave data for
PSMO to  this form yields an exponent $(\nu - \beta)\approx 0.38 \pm
0.05$ [solid line in Fig. 2(c)]. This exponent should be
compared to the value 0.34 predicted for a Heisenberg FM and to the
measured values for iron (0.36 $\pm$ 0.03), cobalt (0.39 $\pm$ 0.05) or
nickel (0.39 $\pm$ 0.04) \cite{collins}. The data for NSMO, on
the other hand, show no evidence of the expected spin wave collapse at
T$_c$,  just as in La$_{0.67}$Ca$_{0.33}$MnO$_3$ \cite{lynn}.
Thus, while the low-temperature spin-wave stiffness for PSMO
and NSMO is $D(0)\approx$ 165 meV-\AA$^2$ (very similar to the 170
meV-\AA$^2$  for La$_{0.67}$Ca$_{0.33}$MnO$_3$ \cite{lynn}),
it is evident that the spin dynamics of these two systems  becomes different as $T
\rightarrow T_C$.

The striking differences between the spin-dynamical behavior for PSMO and
NSMO  as  $T \rightarrow T_C$ are illustrated in  Figure 3.
The data shown in this figure are energy
(constant-q) scans for an arbitrary wavevector q=0.08 reciprocal lattice
units (rlu).  The peaks at the positive side of the energy axis correspond
to neutron energy-loss while those at the negative side are for neutron
energy-gain. A  weak non-magnetic elastic incoherent contribution at
$\hbar\omega=0$  has been subtracted.
Both data sets  show spin wave excitations of
similar energies at the lower temperatures and these energies decrease as
the temperature is increased.
In  PSMO [Fig. 3(a)], the excitations soften and become more
intense as
$T\rightarrow T_C$ as expected for a conventional FM.  With increasing
temperature, a central diffusive component develops.
This central diffusive peak is possibly related to the longitudinal spin fluctuations from a
Heisenberg FM at high temperatures, where the spin-wave description
is not strictly valid \cite {lovesey}, but it may already be a precursor of
the spin dynamics seen in the lower $T_C$ NSMO, which   is definitely
unconventional.  Above $\approx 0.91 T_C$, the spin wave energy
renormalization in NSMO is slower than in PSMO, while the spin wave
intensities damp dramatically and the excitation spectrum is dominated
by the central ($\hbar\omega=0$) component.
This strong diffusive component, which  persists at least
up to $T=1.25T_C$ with little $T$-dependence,  has been attributed to the
presence of magnetic polarons \cite{teresa}.
The magnetic nature of this diffusive component was verified by performing
measurements  close to the (0 0 2)  Brillouin zone center which showed
 the expected intensity drop due to the Mn magnetic form factor.

To characterize the nature of the diffusive component of both
samples, we performed systematic measurements of the
static wavevector-dependent susceptibility $\chi_{\bf q}(\hbar\omega=0)$,
 and the static spin-correlation function. These were ``elastic'' neutron
scattering measurements performed with and without energy
analysis \cite{collins,lovesey}.
The temperature dependence of $\chi_{\bf q}(\hbar\omega=0)$
is shown in Figs. 4(a) and 4(b) for several selected wave vectors.
For the PSMO, the susceptibility shows a sharp peak at the FM transition
as expected for a conventional FM while the data for NSMO
 exhibits a broad peak with a maximum at a temperature somewhat below $T_C$.
The profiles of the  measured static spin-correlation function were
least-squared fitted to an Ornstein-Zernike cross
section \cite{collins} convoluted with the instrumental resolution.  From these fits we
obtained $\kappa(T) = 1 / \xi(T)$, where $\xi(T)$ is the spin-correlation
length.
Figures 4(c) and 4(d) show the temperature dependence of the spin
correlation length $\xi$, and the insets show the corresponding temperature
dependence of $\kappa(T)$ for both systems. For PSMO, the
spin-correlation length $\xi(T)$ increases
on cooling and diverges ($\xi(T)>100$ \AA) close to $T_C$, indicating the
onset of long-range FM order. In contrast,
 the spin-correlation length $\xi$  for NSMO is relatively insensitive to $T$
and remains small ($\approx$ 20 \AA) at $T_C$.
$\xi(T)$ grows to over 100 \AA  \ only at $T \approx 0.95T_C$, consistent
with the presence of magnetic polarons \cite{teresa}.

Our results clearly demonstrate that while the magnetism of these two ferromagnetic
manganites in
the low temperature metallic state may be similar, their spin dynamical behavior around the transition
temperature can be drastically different.  Therefore, it
becomes clear that magnetism alone
cannot explain the exotic spin dynamical properties in these systems and that
the increased electron-lattice coupling plays a dominant role. 
In Table I we emphasize this point by summarizing the results of our experiments 
as well as those for La$_{0.7}$Sr$_{0.3}$MnO$_3$ \cite{martin},
La$_{0.7}$Pb$_{0.3}$MnO$_3$ \cite{toby} and 
La$_{0.7}$Ca$_{0.3}$MnO$_3$ \cite{lynn}.   The various manganites in this table 
have been organized in order of decreasing $T_C$.
For the three highest $T_C$ systems, the values of $D(0)/kT_C$ are characteristic 
of typical localized ferromagnets.  
However, larger $D(0)/kT_C$ values, more characteristic of itinerant ferromagnets \cite{herb}, 
are observed in the last two systems.  This systematic trend occurs despite the 
fact that with decreasing $T_C$, the zero temperature insulating state is approached.

We are grateful to D. Belanger, J. W. Cable, J. F. Cooke,  H. A.
Mook, S. E. Nagler, and  R.M. Moon for helpful discussions. We would also like to acknowledge the expert technical support provided to us by R. G. Maples, S. Moore and G. B. Taylor. Oak Ridge National Laboratory is managed by Lockheed Martin Energy Research Corp.  for the U.S. Department of Energy under contract number DE-AC05-96OR22464.

\begin{figure}
\caption{Temperature dependence of the (0 0 1) Bragg reflection
of (a) PSMO, and (b) NSMO on cooling (full circles) and warming (open circles).
Spin-wave dispersion curves of (c) NSMO (solid circles) and  PSMO (open circles)throughout the zone $ (0, 0, 1 + \zeta)$  at $T=10$K. The
data for PSMO are from Ref. [17].}
\label{autonum}
\end{figure}

\begin{figure}
\caption{Spin-wave energies vs. q$^2$ for (a) PSMO, and (b) NSMO.
(c) Temperature dependence  of the spin-wave stiffness constant $D(T)$ vs.
$T/T_C$ for PSMO (open circles) and NSMO (closed circles). The solid
line is the fit to the mode-mode coupling and hydrodynamic theories at high
temperatures.  The dashed line is an extrapolation to $T=0$ from the
low-temperature mode-mode coupling theory.}
\end{figure}

\begin{figure}
\caption{Energy scans for (a) PSMO, and (b) NSMO at $q=0.08$ rlu using the
same spectrometer setup. }
\end{figure}

\begin{figure}
\caption{(a) Temperature dependence of the static wave vector-dependent
susceptibility  $\chi_{\bf q}(\hbar\omega=0)$ for PSMO at q=0.027,0.033
and 0.039 rlu. (b)   $\chi_{\bf q}(\hbar\omega=0)$  for NSMO at
q=0.025,0.030 and 0.035 rlu. Temperature dependence of the
spin correlation length for (c) PSMO, and (d) NSMO. Insets: temperature
dependence of the inverse spin-correlation length. }
\end{figure}

\begin{table}

\begin{tabular}{clccc}
{Manganite}&{$T_C$ (K)}&{$D(0)$ (meV-\AA$^2$)}&{$D(0)/kT_C$ (\AA$^2$)} \\
\tableline
La$_{0.7}$Sr$_{0.3}$MnO$_3$, Ref. [13]&378&188&5.8 \\
La$_{0.7}$Pb$_{0.3}$MnO$_3$, Ref. [14]&355&134&4.4  \\
Pr$_{0.63}$Sr$_{0.37}$MnO$_3$,  this work&300.9&165&6.4 \\
La$_{0.7}$Ca$_{0.3}$MnO$_3$, Ref. [11]&250&170&7.9&  \\
Nd$_{0.7}$Sr$_{0.3}$MnO$_3$,  this work&197.9&165&9.7 \\
\end{tabular}

\caption{Summary of magnetic properties of various perovskite manganites  $A_{1-x}B_x$MnO$_3$ ($x \approx 0.3$) in order of decreasing $T_C$.}

\label{Table I}

\end{table}

\end{document}